\documentclass[twocolumn,showpacs,superscriptaddress,pre,amsmath]{revtex4}
\usepackage{graphicx} % Include figure files
\usepackage{dcolumn}  % Align table columns on decimal point
\usepackage{bm}       % bold math
\usepackage{amssymb}

\begin{document}

%\sloppy
%\draft
\title{Universal behavior in 
populations composed of excitable and self-oscillatory elements}
%\title{Generalized aging transition in oscillatory networks}

\author{Diego Paz\'o}
\altaffiliation{Current address: Instituto de F\'{\i}sica de Cantabria 
(CSIC-UC), E-39005 Santander, Spain.}
%\email{pazo@mpipks-dresden.mpg.de}
\affiliation{Max-Planck-Institut f\"ur Physik komplexer Systeme, 
N\"othnitzer Stra{\ss}e 38, 01187 Dresden, Germany}

\author{Ernest Montbri\'o}
%\email{ernest@imedea.uib.es}
\affiliation{Departament de F\'{i}sica, Universitat de les Illes Balears, E-07122 Palma de Mallorca, Spain}
\affiliation{Institut Mediterrani d'Estudis Avan\c{c}ats,
IMEDEA (CSIC-UIB), E-07122 Palma de Mallorca, Spain}
\affiliation{Computational Neuroscience, Technology Dept., Universitat Pompeu Fabra, 08003 Barcelona, Spain}
\date{\today}

\begin{abstract}

We study the robustness of self-sustained oscillatory activity in a globally coupled ensemble 
of excitable and oscillatory units. The critical balance to achieve collective self-sustained oscillations is analytically established. 
We also report a universal scaling function for the ensemble's mean 
frequency. Our results extend the framework 
of the `Aging Transition' [Phys.~Rev.~Lett.~{\bf 93}, 104101 (2004)]
including a broad class of dynamical systems potentially relevant in biology.

\end{abstract}


\maketitle

Large networks of interacting dissipative systems are appropriately modeled in terms of coupled nonlinear differential equations, which successfully reproduce a huge variety of the dynamical patterns found in nature. 
In many cases, the individual systems present time-periodic behavior 
and may achieve
a certain degree of global synchronization despite the unavoidable 
differences among them~\cite{winfree,Kur84}. This subject
has 
attracted a great deal of both theoretical 
and experimental interest during the last decades \cite{PRK01}.  

However, 
the robustness of the macroscopic synchronized oscillations 
in a mixed population of self-oscillatory and non-self-oscillatory elements 
has only been addressed very recently by Daido and Nakanishi~\cite{daido2004}. 
Interestingly, they report a general scenario, called `Aging Transition',
characterized by a universal (i.e. independent of the oscillator type) scaling function. 
In~\cite{daido2004} the Aging Transition was found 
in populations
of oscillators in which some of them lose
their self-oscillatory activity (by deterioration or `aging')  
through an inverse Hopf bifurcation.

In this Letter 
we present a extension of the Aging Transition taking place in a new 
class of systems in which the oscillatory behavior is lost in a saddle-node (SN) bifurcation. Such systems are of particular relevance since their resulting dynamics is excitable and thus 
of interest in many areas of physics, chemistry and biology~\cite{meron92,winfree,Keener,HI98}.

For instance, 
a remarkable example of macroscopic synchronization 
is found among the pacemaker cells in the sino-atrial node which initiate the heartbeat. 
When a certain ratio of cells are damaged by disease, the lack of an
adequate synchronized state requires the implantation of an electronic pacemaker.
Nevertheless, new techniques aiming to create biological pacemakers 
have been recently proposed  as a very promising 
alternative  to electronic ones~\cite{biopacemaker2}. 
One of them consists in the creation of an aggregate 
of biological pacemaker cells in some region(s) of the ventricle, 
converting excitable heart cells into pacemaker cells by gene transfer~\cite{biopacemaker},
in an ``inverse-aging'' transition. 
This is feasible since the heart's pacemaker and excitable cells are 
very similar (they are both identical pacemaker cells in the early embryonic heart, 
and differentiate as the development progresses).  
In this context, the robustness of the aggregate's self-oscillating 
activity in a mixed population of self-oscillatory (converted) and excitable (unaltered)
elements seems to be particularly relevant.

\begin{figure}
\includegraphics[width=3.in]{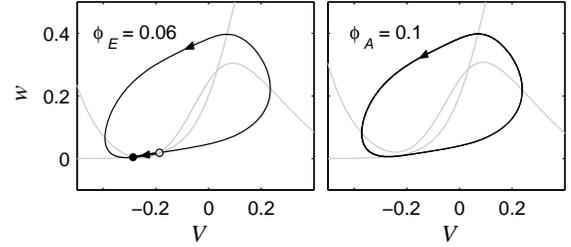}
\caption[]{Morris-Lecar equations (\ref{ml}) in the excitable (left panel) and in the oscillatory/pacemaker (right panel) regimes. 
Both dynamical states are linked through a saddle-node bifurcation on invariant circle (SNIC) 
at  $\phi^* \simeq 0.076$ [see Eq.~(\ref{ml1})]. 
The nullclines $\dot V =0 $ and $\dot w=0$ are depicted with gray lines.}
\label{fig_snic}
\end{figure}

In our study, we have considered systems that cease their oscillatory behavior
through a SN bifurcation on invariant circle (SNIC)~\cite{HI98}, 
also called SN homoclinic bifurcation~\cite{Kuznetsov} (see Fig.~\ref{fig_snic}). 
This is the simplest possible scenario 
linking excitable and oscillatory dynamics,
and there is only one attractor at each parameter value.
The excitable regime (at one side of the bifurcation) is very well known 
in theoretical neuroscience, where it is referred to as 
Class I excitability~\cite{rinzel89}.

Additionally, we make the following assumptions: 
\begin{itemize}
\item[(i)]The population (of size $N$) is divided into two groups, 
consisting of
$pN$ [$1/N  \le p \le (N-1)/N$] identical excitable units ($S_E$), and of
$(1-p)N$ identical active 
units ($S_A$). The elements in the population are ordered according 
to an index $j$, such that $S_A \equiv j \in \{1,\ldots, (1-p)N\}$ 
and $S_E \equiv j \in \{(1-p)N+1,\ldots, N\}$.
\item[(ii)] 
A linear all-to-all coupling is assumed, for the state variable $\mathbf{x}_{\mathit{j}}$
of each unit: 
\begin{equation}
\mathbf{\dot x}_{\mathit{j}}  = \mathbf{F}_j(\mathbf{x}_j)+\frac{K}{N} \sum_{k=1}^{N}(\mathbf{x}_k-\mathbf{x}_j),
\label{general}
\end{equation}
where $\mathbf{F}_j=\mathbf{F}_A$ for $j\in S_A$ and 
$\mathbf{F}_j=\mathbf{F}_E$ for $j\in S_E$.  
In general, the coupling term in (\ref{general}) could simply enter through a single variable of $\mathbf{x}_j$ (e.g.~the membrane potential, in a cell model). 
This has in many cases the same synchronizing effect (but see~\cite{han}).
\end{itemize}

\begin{figure}
\includegraphics[width=3.in]{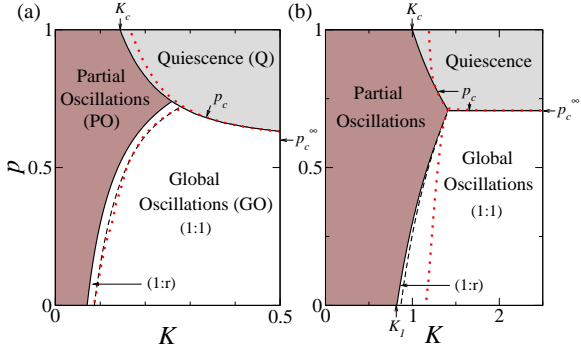}
\caption[]{(color online) $(K,p)$ phase diagram for a mixed population
of oscillatory and excitable elements: (a) Morris-Lecar (\ref{ml}) with  $\phi_A=0.1$ and $\phi_E=0.06$; (b) phase model (\ref{phase_pop}) with $b_A=0$ (oscillatory) and $b_E=\sqrt{2}$ (excitable). 
Inside the region of Global Oscillations the black dashed line separates $1:1$ (right) from other 
$1:r$, $0<r<1$ (left) frequency ratios, within the region GO.
The red dotted curves are the deformation of the solid lines when heterogeneity 
in each subpopulation is considered (see text): 
$\gamma=$ (a) 0.01 and (b) $0.6$.} 
\label{diagrams}
\end{figure}

First we introduce our findings using an ensemble of globally coupled 
Morris-Lecar (ML) units~\cite{morris81}. The ML equations were
obtained from the study of
the electrical activity of the barnacle muscle fiber, and later  
they were
popularized 
as a reduced 
model of
excitability~\cite{rinzel89}. 
We have used the adimensionalized version proposed 
in~\cite{er_kop90}, with the coupling term (proportional to $K$) entering through
the voltage variable. The system is
\begin{subequations}
\label{ml}
\begin{eqnarray}
 C \dot V_j&=& g_L (-V_L-V_j)+g_{Ca} m_\infty(V_j)(V_{Ca}-V_j)   \label{ml1}\\
&-& g_K w_j(V_K+V_j)+\phi_j(0.2-V_j)+ \frac{K}{N}\sum_{i=1}^N(V_i-V_j),  \nonumber\\
\dot w_j&=&  \lambda(V_j)(w_\infty(V_j)-w_j), \label{ml2}
\end{eqnarray}
\end{subequations} 
where $m_\infty(V_j)=0.5\{1+\tanh[(V_j-v_1)/v_2]\}$,
$w_\infty(V_j)=0.5\{1+\tanh[(V_j-v_3)/v_4]\}$, and
$\lambda(V_j)=\lambda_0\{1+\cosh[(V_j-v_3)/v_4]\}$.
Several constants~\footnote{$g_L=0.5$, $V_L=0.4$, $C=g_{Ca}=V_{Ca}=1$, $g_K=2$, 
$V_K=0.7$, $v_{1,2,3,4}=(-0.01,0.15,0.1,0.145)$, $\lambda_0=0.33$.} were taken from~\cite{er_kop90} with the external current 
set equal to zero. 
The term $\phi_j(0.2-V_j)$ in Eq.~(\ref{ml1}) controls the dynamics of
an isolated element~\footnote{It has been introduced to account 
in the simplest way  for the pacemaker (dubbed ``funny'') 
current present in the cardiac pacemaker cells. This current 
is active for small membrane potential and vanishes just
after the action potential upstroke ($V\sim 0.2$).}.
In our model $\phi_j$ 
takes only two possible values $\phi_A$ and $\phi_E$.
For $\phi_A>\phi^* \simeq 0.076$ the dynamics of an 
isolated cell is self-oscillatory, 
whereas for $\phi_E<\phi^*$ 
it 
is excitable. 
Both behaviors are linked by a SNIC at $\phi^*$ (Fig.~\ref{fig_snic}).

The system (\ref{ml}) is described in terms of two control parameters, the 
ratio $p$ of excitable units, and the coupling strength $K$. 
Figure \ref{diagrams}(a) shows the $(K,p)$ phase diagram 
with the dynamical regimes found in the ensemble of ML units. 
For small coupling $K$, each element in the population essentially 
maintains its intrinsic dynamics. 
Such state is labeled as {\em Partial Oscillations (PO)} because 
the collective state is not entirely oscillatory (the excitable elements are 
not able to oscillate, 
they only exhibit small amplitude ``pulsations''). 
For large values of the coupling strength $K$, all the elements in the ensemble exhibit the 
same 
behavior, but depending on the ratio $p$ they are all at rest 
({\em Quiescence (Q)}) or all oscillating ({\em Global Oscillations (GO)}). 
The region 
{\em GO} contains a thin stripe-shaped subregion, 
where all the units are oscillating, but with frequency ratios different from $1:1$.

The structure of the parameter space in Fig.~\ref{diagrams}(a) is reproduced for
other oscillator types like the one proposed by Eguia {\em et al.} \cite{eguia}
and a phase model close to a SNIC (so-called `active rotator'~\cite{shinomoto_kuramoto}).
The equations for the globally-coupled phase model are
\begin{equation}
\dot \theta_j=1- b_j \sin\theta_j \\ +\frac{K}{N} \sum_{l=1}^{N} \sin(\theta_l - \theta_j),
\label{phase_pop}
\end{equation} 
where the parameter $b_j$ characterizes the non-uniformity of the phase rotation. 
Figure~\ref{diagrams}(b) shows 
that the simple phase model (\ref{phase_pop}) 
qualitatively behaves as the ensemble of ML oscillators [Fig.~\ref{diagrams}(a)].
The only significant difference 
is the shape of the region inside {\em GO}
with different rotation numbers in each subpopulation: wedge-shaped for the phase model and stripe-shaped for  
``complete'' models. 
For phase models (other implementations were tested),
leaving the {\em Q} region implies generically a transition 
to {\em GO} with $1:1$ frequency ratio, or to {\em PO} 
($1:0$). For 
coupled oscillators,
the transition from the quiescent state to oscillations
with other frequency ratios ($1:r$) 
is of codimension one. From our simulations with the ML equations
we have indications that in this transition
the structure of phase space is a Cherry flow (see e.g.~\cite{maistrenko}), 
although this seems difficult to prove.

In order to check the robustness of the diagrams in Fig.~\ref{diagrams} 
we performed numerical simulations ($N=1000$) considering 
a certain degree of heterogeneity in each subpopulation (see red dotted lines in Fig.~\ref{diagrams}). 
This has been carried out distributing the systems' parameters uniformly around 
$\phi_{A,E}$ (ML) and $b_{A,E}$ (phase model) with a finite width $\gamma$. 
Heterogeneity shrinks the region {\em Q}, since there is need of more coupling strength to compensate the major diversity in the population. As a last remark,
we just notice that the dynamics in the {\em PO} and {\em GO} regions become 
more complex (some regimes cannot be labeled by just one frequency ratio).

The rest of this Letter is devoted to the 
analysis of Fig.~\ref{diagrams} in the absence of heterogeneity (i.e. $\gamma=0$). We will mainly stress those results whose
validity is general, i.e.~irrespective of the particular
dynamical system considered. 
One empirical fact that simplifies the analysis is
that in the whole $K-p$ plane 
%($K\ne 0$) 
all the identical elements are at the same 
state: $\mathbf{x}_{j\in S_A}=\mathbf{x}_A, \mathbf{x}_{j\in S_E}=\mathbf{x}_E$.      
Thus, it is possible to study the dynamics of the population in terms of two asymmetrically
coupled elements:
\begin{subequations}
\begin{eqnarray}
\mathbf{\dot x}_A &=& \mathbf{F}_A(\mathbf{x}_A) + Kp(\mathbf{x}_E-\mathbf{x}_A), \label{2oi}\\
\mathbf{\dot x}_E &=& \mathbf{F}_E(\mathbf{x}_E) + K(1-p) (\mathbf{x}_A-\mathbf{x}_E), \label{2oii} 
\end{eqnarray}  
\label{2o}
\end{subequations} 
with $\mathbf{x}=(V,w)$ for the 
ML cells, and $\mathbf{x}=\theta$ for the phase model. 
This simplification allows to study the $N\rightarrow \infty$ limit 
and to adjust $p$ continuously. In fact, reduction (\ref{2o}) was used to 
efficiently compute Fig.~\ref{diagrams}.
And in particular for the phase model (\ref{phase_pop}), the bifurcation lines limiting 
{\em Q} can be easily calculated analytically {for $b_A=0$}:
$p_c=1/K$ and $p_c=1/b_E$,
with a degenerate point at $(K,p)=(b_E,b_E^{-1})$. Other lines must be computed
numerically~\footnote{The point $K_1=0.8205\ldots$ where the
line limiting {\em PO} touches the axis $p=0$ 
can be approximated by an expansion
in powers of $\beta \equiv b_E-1$: 
$K_1^2=4\beta/3 + {\cal O} (\beta^2)$. Taking
$\theta_A=t$, from the non-autonomous ODE:
$\dot \theta_E= 1 -b_E\sin{\theta_E}+ K\sin(t-\theta_E)$,
$K_1$ can be estimated thanks to the symmetries 
of the saddle-node solution.}.

An important feature of the phase diagrams in Fig.~\ref{diagrams} is
the asymptotic value of the bifurcation line limiting {\em Q}: 
$p_c^\infty \equiv p_c(K\rightarrow\infty)$. 
Remarkably, $p_c^\infty$ can be analytically estimated in a simple way. 
The calculation is based on the fact that 
close to $p_c$ the dynamics evolves for long time inside
a (slow) region where the SN bifurcation takes place.
Therefore, resorting to the normal form of a SN bifurcation, 
we obtain (after rescaling space and time):
\begin{subequations}
\begin{eqnarray}
\dot z_A &=& a \epsilon_A - z_A^2+{ K}p(z_E-z_A), \label{sn1}\\
\dot z_E &=& a \epsilon_E - z_E^2+{ K}(1-p)(z_A-z_E), \label{sn2}
\end{eqnarray} 
\label{sn}
\end{subequations} 
with $\epsilon_A<0$ and $\epsilon_E>0$ according to the oscillatory (no fixed point) and the 
excitable (one stable and one saddle fixed point) regimes
at both sides of the SN bifurcation (at $\epsilon=0$). 
In the limit $K \rightarrow \infty$,
the quadratic terms can be neglected at the bifurcation point. From the nullclines of
Eq.~(\ref{sn}) [$z_E=z_A-(a\epsilon_A-z_A^2)/Kp$ and $z_A=z_E-(a\epsilon_E-z_E^2)/K(1-p)$] we obtain
\begin{equation}
p_c^\infty=\frac{\epsilon_A}{\epsilon_A-\epsilon_E}.
\label{pc}
\end{equation}
This is one of the main results of the present work. Equation (\ref{pc}) gives 
the critical proportion of 
excitable 
and oscillatory elements in order for
the whole population to become self-oscillatory. 
Only three parameters are needed: the value of the control parameter at the 
SN bifurcation, and the distances, $\epsilon_A$ and $\epsilon_E$, 
of the oscillatory and 
excitable 
elements to the bifurcation point. Specifically, for the ML model we have
$\epsilon_{A,E} \simeq (\phi^*-\phi_{A,E})$,
that according to (\ref{pc}) yields 
$p_c^\infty \simeq 0.60$, in good agreement with the 
numerical results (see also Fig.~\ref{fig_ml} below) \footnote{Eq.~(\ref{pc}) has been validated
with a population of oscillators of the type  
proposed in~\cite{eguia}.}. 
Remarkably we find that Eq.~(\ref{pc}) holds also for an ensemble 
of dynamical systems at both sides of a {\em Hopf} bifurcation (we skip here the proof, but
cf.~Eq.~(4) in \cite{daido2004} for a particular case).

Our next results concern 
the behavior of the mean ensemble's frequency, 
which is a natural measure for the global oscillatory activity. 
The ensemble's average of the individual frequencies is  
\begin{equation}
\Omega=\frac{1}{N}\sum_{j=1}^N \Omega_j ,
\label{O}
\end{equation} 
where $\Omega_j$ represents the mean frequency of the $j$-th element. 
Obviously, under the assumption in Eq.~(\ref{2o}),
$\Omega=(1-p)\Omega_A+p\Omega_E$  (with $\Omega_A=\Omega_{j \in S_A}$,
$\Omega_E=\Omega_{j \in S_E}$).
Fig.~\ref{fig_ml} shows the 
average frequency (\ref{O})---normalized by the frequency at $p=0$: $\tilde\Omega(p)
\equiv \Omega(p)/\Omega(0)$---,  
and the  individual frequencies $\Omega_j(p)$ (right panels) 
for different values of 
$K$. 
Notice that for small coupling the transition to 
{\em Quiescence} ($\tilde\Omega=0$) occurs trivially at $p_c=1$. 
Such behavior changes above $K_c\simeq 0.144$, 
where the transition begins to take place at $p_c<1$. 
Also, note that for intermediate values of $K$ 
the curves present a 
steplike profile, 
corresponding to the stripe-shaped region in Fig.~\ref{diagrams}(a). 
This can be seen in Figs.~\ref{fig_ml}(c,d) where, in the corresponding range of $p$, 
the two observed frequencies in the ensemble are not $1:1$ (neither $1:0$) related.

\begin{figure}
\includegraphics[width=3.2in]{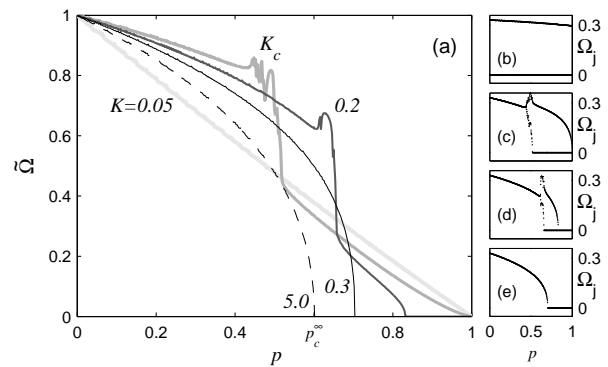}
\caption[]{(a) Normalized ensemble's frequency 
$\tilde \Omega(p)$
%$\tilde \Omega(p)\equiv \Omega(p)/\Omega(0)$ 
for different values of the coupling 
strength 
$K$ in a population of oscillatory and excitable ML units ($N=500$) . 
The dashed line for large coupling ($K=5$) shows the convergence of $p_c$
to the value predicted by Eq.~(\ref{pc}).
Panels (b-e) show the individual frequencies for: $K=0.05$ (\emph{PO})(top), 
$K_c \simeq 0.144$, $K=0.2$, and $K=0.3$ (transition  \emph{GO-Q}) (bottom).} 
\label{fig_ml}
\end{figure}

The quantity $\Omega$ presents 
interesting universal properties 
around $K_c$.
Assuming that Eq.~(\ref{2o}) 
exhibits a SN bifurcation at some $p_{sn}(K)$ we have : 
$p_c=1<p_{sn}(K)$ (for $K<K_c$), $p_c=1=p_{sn}(K)$ (at  $K=K_c$) and $p_c=p_{sn}(K)<1$ (for $K>K_c$). 
And noting that
\begin{enumerate}
\item[(i)] Only the active (self-oscillating) elements contribute to $\Omega$: $\Omega=(1-p)\Omega_A$.
[$\Omega_E=0$].
\item[(ii)] In a SNIC, the frequency of the oscillating elements scales as a square root:
$\Omega_A \simeq h (p_{sn}-p)^{1/2}$. 
\end{enumerate} 
we find 
that $\Omega$ grows from zero as a power of the distance to 
the critical $p_c$:
\begin{equation}
\Omega=(1-p)h(p_{sn}-p)^{1/2}\propto(p_c-p)^\beta,
\label{beta}
\end{equation} 
with three different values of the exponent $\beta$:
$1$ (for $K<K_c$, $p_{sn}>p_c=1$), $3/2$ (at $K=K_c$, $p_{sn}=p_c=1$), $1/2$ (for $K >  K_c$,
$p_{sn}=p_c<1$). 

\begin{figure}
\includegraphics[width=2.8in]{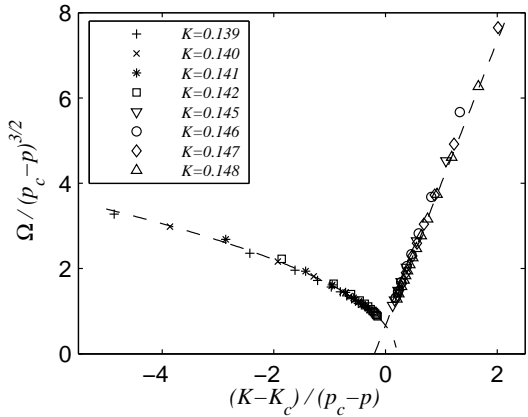}
\caption[]{Fitting to Eq.~(\ref{scaling_daido}) for an ensemble of $N=1000$ ML 
elements, $K_c=0.14386(0)$. Dashed lines stem from Eq.~(\ref{scaling_daido}) 
with fitting parameters $g=5.01$ and  $h=0.66$.} 
\label{fig_daido}
\end{figure}

In a neighborhood of $(K=K_c, p=1)$, we may obtain a
universal scaling function by assuming that
the shift of $p_{sn}$ is approximately linear on $K$: $p_{sn} \simeq 1 +g(K_c-K)$.
Recalling that $p_c=1$ for $K\le K_c$, we obtain from Eq.~(\ref{beta}) the expressions
\begin{eqnarray}
\left\lbrace 
\begin{array}{l}
\Omega = h (p_c-p) [p_c-p+g (K_c-K)]^{1/2},  \,       (K \le K_c) \\ 
\Omega = h (1-p) (p_c-p)^{1/2}, \qquad (K>K_c)
\end{array} 
\right. 
\label{2scal}
\end{eqnarray}
These distinct scalings 
may be condensed into the single scaling function (by approximating
$1 \simeq  p_{c} - g(K_c-K)$ in the second equation):
\begin{equation}
\Omega= h (p_c-p)^{3/2} \Phi\left( g\frac{K-K_c}{p_c-p} \right),
\label{scaling_daido}
\end{equation} 
with $\Phi(x)$ $[=\sqrt{1-x}$ \,($x<0$)], $[=1+x\,  (x>0)]$. 
We obtain then two scaling regions with common fitting parameters $g$ and $h$,
see Fig.~\ref{fig_daido}. 

It is important to note that the  scaling in Fig.~\ref{fig_daido} coincides with the one reported in~\cite{daido2004} for the {\em amplitude} of oscillations in a mixed population of oscillators close to a Hopf bifurcation. Such coincidence lies in the same
asymptotic dependence (square-root law) for the cycle's amplitude in the case of a Hopf bifurcation 
and for the cycle's frequency in the case of a SNIC.

In conclusion, we have demonstrated that for systems close to a SNIC, 
a transition to global quiescence occurs as the ratio $p$ of 
excitable elements in the ensemble exceeds a certain value $p_c$.
Such transition is a generalization of the Aging Transition 
reported in \cite{daido2004} and is  characterized by a universal 
scaling function relating the mean frequency 
$\Omega$ with $p$ and the coupling strength $K$. 
Additionally, we derive an analytical estimation for $p_c$ in the large $K$ limit %($p_c^\infty$) 
which holds for both Hopf- and SNIC-mediated Aging Transitions.

Our results might be of importance in several situations, 
since excitability is a typical feature of many physical, chemical and biological systems. 
Hence, our work is a first step in modeling the competition between
excitable and oscillatory dynamics, with possible extensions to extended media and complex networks. 
Finally, our findings could also be relevant in populations of excitable systems when some 
of the elements turn self-oscillatory due to the presence of noise (coherence resonance~\cite{coh_res}), a situation 
common in neuroscience.

%\acknowledgments
Helpful comments from M.~A.~Mat\'{\i}as are gratefully acknowledged.
We also thank C.~Monterola and G.~Kalosakas for a critical reading of the manuscript.

\end{document}